\documentstyle[preprint,eqsecnum,aps,epsfig]{revtex}
\begin{document}

 \tightenlines

\def\pa{\partial}
\def\ov{\over}
\def\non{\nonumber }
\def\beq{\begin{equation} }
\def\eeq{\end{equation} }
\def\beqa{\begin{eqnarray}}
\def\eeqa{\end{eqnarray}}

\title{Chiral Soliton Model vs Pentaquark Structure for $\Theta(1540)$}
\author{R. Ramachandran}

\address{~The Inter University Centre for Astronomy and Astrophysics, University of Pune Campus, Pune, 411007, India\\ and}
\address{~Department of Physics, University of Pune Campus, Pune, 411007, India  \\
{\tt rr@iucaa.ernet.in, rr@physics.unipune.ernet.in}}

\maketitle

\vskip1.5cm

\begin{abstract}

The exotic baryon $\Theta^+$(1540 MeV) is visualized as an expected (iso)
rotational excitation in the Chiral Soliton Model. It is also argued as a
Pentaquark baryon state in a constituent quark model with strong diquark
correlations. I contrast the two points of view; observe the similarities and
differences between the two pictures. Collective excitation, characteristic of
Chiral Soliton Model points toward small mixing of representations in the wake
of $SU(3)$ breaking. In contrast, Constituent quark Models prefer near
``ideal'' mixing, similar to $\omega - \phi$ mixing.

\end{abstract}
\vskip0.5cm

\section{Introduction}

An exotic baryon $\Theta^+(1540 MeV)$ with the quantum numbers of $K^+n$ has
been observed as a very narrow width state by several groups\cite{Spring}. It
has hypercharge $Y = 2$, the third component of isospin $I_3 = 0$. Since such
a state has not been seen so far in the $K^+p$ channel, $I=1$ is ruled out and
so $\Theta^+ $ is an isosinglet. While this is
yet to be confirmed by some other experimental groups \cite{negresult} that  don't see
evidence for a narrow state so far, there is a consensus
that there is enough evidence to warrant its inclusion in the 2004 edition of
Particle Data Book \cite{pdg}. 
The minimal $SU(3)$ assignment for such a state is at the top ($Y=2,I=0$) of
$\{\overline{10}\}_F$ representation with $Y=1\ I={1\ov 2}\ (N^0,N^+)$, $Y=0\ I=1
\ (\Sigma^-,\Sigma^0,\Sigma^+)$ and $Y=-1\ I={3\ov 2}\ 
(\Xi^{--},\Xi^-,\Xi^0,\Xi^+)$ as other members of the family. 

The term exotic refers to the fact that such a state is not realized as usual
three-quark composite, since positive strangeness for the baryon calls for an
$\overline{s}$ quark in it. Minimal quark configuration is $udud\overline{s}$. Even though the spin and parity of
$\Theta^+(1540)$ are yet to be determined experimentally, most of the
theoretical analysis has
carried a general prejudice that it is $J^P = {1 \ov 2}^+$ state.

The static (low energy domain) properties of the baryons are not easily
derived from the underlying QCD on account of the non-perturbative features of
the theory. It becomes necessary to look for other models that are inspired by
QCD to throw further light on the structure and properties of hadrons (mesons
and baryons) that are indeed  color singlet composites of quarks and
antiquarks. Even though the color degrees of freedom are hidden, it is useful to talk about the quark, anti-quark and
gluon content of hadrons as revealed to an external electromagnetic or weak probe in a deep inelastic scattering by 
leptons ($e^\pm,\mu^\pm,\nu, \overline{\nu}$) or photons. These hadron structure
functions (or more correctly their evolution as a function of resolution scale) are
accessible to perturbative QCD. For other non-perturbative properties, one
resorts to study QCD either on a lattice or use other effective theories,
presumably 
derivable from QCD. Chiral Lagrangian Dynamics is one such formalism in which
QCD is seen to express its global (flavour) symmetries through the
pseudo scalar meson degrees of freedom in the large $N_c$ (where $N_c$ is number of
colors) limit. Chiral Lagrangian with the octet of pseudoscalar mesons as
primary fields admits a solitonic mode (Skyrmion)\cite{skyrme}, which provides the baryon
sector. While the ground state in this sector is an $SU(3)$ octet of baryons,
of which the nucleon is the $Y=1\ I={1\ov 2}$ member, other excited states
are rotational (in ordinary and internal $SU(3)$ flavour space) excitations.
Quantization in the collective co-ordinates associated with the Skyrmion
solution gives the spectrum of baryonic states.  

It is possible to show that$ \{\overline{10}\}_F$ baryons, of which the $\Theta(1540)$
is a member as the next rotational excitation after the ground state $\{8\}$,
  which has the nucleon and the well known $\{10\}$ representation of which
  $\Delta$, the 
    isospin quartet at 1232 Mev, is a prominent member. Indeed, Diakanov,
    Petrov and Polyakov \cite{diakanov} predicted the mass and the narrow width of the observed
   $ \Theta^+$ on the basis of the chiral soliton model. There is now a vast
    literature accumulated on the subject treating $\Theta^+$ in terms of the
    chiral soliton model on the one hand \cite{chiraltheta} and in terms of a
    constituent pentaquark model (with special correlations) on the
    other\cite{cqmtheta}. While both
    models can account for the presence of $\{\overline{10}\}_F$ states, they differ
    on what else is expected and what may be the consequence of $SU(3)_F$
    symmetry breaking. We provide the similarities and contrasts of both
    points of view.  There are also efforts to find pentaquark states in
    Lattice QCD\cite{lattice}, which at the moment remains inconclusive. 

\vskip0.5cm
\section{Chiral Soliton Model}

\vskip0.3cm
Effective Lagrangian embodies the chiral symmetry and is a function of
$U(x)$, a unitary $3\times3$ matrix and $\pa_{\mu} U(x)$. The pseudo scalar octet of
mesons are expressed through $U(x) \equiv exp\ ({i \ov F_\pi}
\lambda^a\phi^a(x)),\  a=1,2,..8$;\  $\lambda^a$ are Gell Mann matrices and $x$
denotes $(\vec{x}, t)$. $F_\pi$ is pion decay constant and provides a scale for the masses in the theory. The theory admits finite energy static solutions (for
the classical equations of motion), that has  the form:
$$
U(x) = U_0(\vec{x})=\pmatrix { 
exp \ i f(r) \vec{\tau}\cdot \hat{x} & 0 \cr 
0  & 1 }
$$
The constraint that 
$$
f(r)\to \matrix {\pi\ &\  \mbox{when } r\to 0 \cr
 0 &\ \mbox{when } r\to \infty},
$$ 
ensures that $U(x)$ is both well defined at the origin and can lead to finite
energy configuration. Precise form of $f(r)$ can be obtained numerically from the radial
equation of motion. Quantization is then carried out by identifying the
collective co-ordinates, a set of parameters that label the transformations
that will leave the classical solution invariant. Apart from the centre of mass
of the Skyrmion, the translation of which will indeed yield the overall momentum of the state and
hence the kinetic energy, rotations in space as well as internal space are the other collective 
co-ordinates. To get their dynamics, it is convenient to identify the collective
co-ordinates ${\mathcal {A}}$ by defining 
\beq
U(\vec{x},t) = {\mathcal A}(t)U_0(\vec{x}){\mathcal A}^{-1}(t); \quad {\mathcal A}\in  SU(3)
\eeq
Notice that the solution is left invariant under ${\mathcal A}\to L{\mathcal A}$; $L\in SU(3)$
denoting a $SU(3)$ flavour transformation and under ${\mathcal A} \to {\mathcal A}R$; $R \in SU(2)$, an
element of rotation in space. ${\mathcal A}(t)$ constitute the relevant collective
co-ordinates that embody the rotational and iso-rotational degrees of freedom
for the Skyrmion. The wavefunction for the baryons in various allowed irreducible
representations of $SU(3)$ are given by the equivalent of Wigner D-
functions ${\mathcal{D}}^{\{R\}}_{\alpha,\beta}({\mathcal A})$, where $\{R\}$ stands for
the $SU(3)_F$ representation of the state; $\alpha = (I,I_3\ Y)$ and $\beta =
(I'=J,I'_3=J_3 \ Y'=1)$ respectively denote the  
iso-rotational (isospin $I$,$I_3$ and hypercharge $Y$) and rotational (angular momentum $J,J_3$) state of the baryon.\cite{smr} \cite{guadagnini}.  The relevant Hamiltonian has the form:
\beq
H=M_0+ {1 \ov 2{\mathcal{I}}_1} \sum_{a=1}^{3}J_a^2+{1 \ov 2{\mathcal{I}}_2} \sum_{a=4}^{7}J_a^2 +
(1/N_c \ \mbox{corrections})
\eeq
where ${\mathcal{I}}_{1,2}$ are `moments of inertia' for the rigid rotator model for the
baryon. The interlocking of the spin and isospin is an essential feature of
the soliton sector and generates a constraint on the states that is further
made precise from the presence of the Wess- Zumino term in the effective
Lagrangian for $SU(n_F),\ n_F > 3$. For the $SU(2)$ Skyrmion this is reflected by the fact that the
allowed baryons
 have their $I = J$. For $SU(3)$, the constraint admits in the spectrum
only those $SU(3)$ representations that have $Y = 1$ member and the spin $J$ for
the set will be the same as the isospin values of all the $Y = 1$ members of
the representation. For $n_F > 3$, the angular momentum of the state assumes value(s) of isospin of the $Y=1$, $SU(n_F-2)$ singlet(s)\cite{manohar}.  

Since every $SU(3)$ unitary irreducible representation is given by a pair of
indices $(p,q)$ (the wave function has $p$ indices that transform
like $\{3\}$ and $q$ indices like $\{\overline{3}\}$) and the second casimir operator
$C_2(p,q) \equiv \sum_{a=1}^{8} J_a^2 = {1 \ov 3} (p^2+q^2+pq+3(p+q))$ and
$J_8 = -{\sqrt{3} \ov 2} Y$, we can read off the mass spectrum (Wess-Zumino constraint implies $|p-q| =0\ \mbox{modulo}\ 3$) as:

$$
E^J(p,q) =M_0 +{1 \ov 2{\mathcal{I}}_2} C_2(p,q) + ({1 \ov 2{\mathcal{I}}_1}-{1 \ov 2{\mathcal{I}}_2})J(J+1)-
{3 \ov 8{\mathcal{I}}_2}.
$$

If ${\mathcal{I}}_1 > {\mathcal{I}}_2$, we will have the observed sequence of states; ground state
$\{8\}_{1/2}$ followed by $\{10\}_{3/2}$ and the just discovered
$\{\overline{10}\}_{1/2}$. From the central or average values of the masses in these 
multiplets, $1115$ MeV for $ \{8\}_{1/2}$, $1382$ MeV for $\{10\}_{3/2}$ and $1755$ MeV for
$\{\overline{10}\}_{1/2}$, we may determine values of ${\mathcal {I}}_{1,2}$
and get the sequence of further excitations to be at(in MeV)
1784 $\{27\}_{3/2}$, 1967 $\{35\}_{5/2}$, 2155 $\{27\}_{1/2}$, 2570
$\{64\}_{5/2}$, 2588 $\{35\}_{3/2}$, 2707 $\{81\}_{7/2}$, 2959
$\{\overline{35}\}_{1/2}$ and so on. 

An important observation made on the basis of the chiral soliton picture has
to do with the narrow width of the baryons in $\{\overline{10}\}_{1/2}$. Diakanov
{\it et al} \cite{diakanov} attribute this to the next to leading order (in $1/N_c$)
correction for the meson baryon couplings. While the leading order for all
transitions in the soliton sector is characterized by $G_0$, there are two
further terms with strengths $G_1$ and $G_2$ in the next order. For example,
the $(B_8B_8M_8)$ has two distinct $SU(3)$ symmetric couplings $D$ and  $F$,
usually given instead in terms of $g_{\pi NN} (= D +F)$ and $\alpha (= D/(D+F))$
that are expressed through  
\beqa
g_{\pi NN} &=& {7 \ov 10} (G_0 + {1 \ov 2}G_1 + {1 \ov 14}G_2)\\
\alpha &=& {9 \ov 14} {G_0 + {1 \ov 2}G_1 - {1 \ov 6}G_2 \ov G_0 +
    {1 \ov 2}G_1+ G_2} 
\eeqa

In view of the fact that the experimental value of $\alpha = 0.65$ is very close to 9/14, we expect $G_2$ to be small and negligible. 

The decouplet baryon  decay couplings ($B_{10}B_{8}M_{8})$ are characterized
by the factor $G_0 + {1 \ov 2}G_1$ and the antidecouplet decays
($B_{\overline{10}}B_{8}M_{8}$) are governed by the  term $G_0 - G_1
-{1 \ov 2}G_2$. Diakanov estimates $G_1/G_0$ to be in the range 0.4 to 0.6
with the result $G_{\overline{10}}/G_{10} \sim 1/3\ \mbox{to}\  1/5$. Indeed, with a
bit of adjustment in the value of $G_1$ and $G_0$, considerable suppression
for width of antidecouplet baryons can be realized. 

When $SU(3)$ is explicitly broken there are two further consequences. There
will be splitting within each $SU(3)$ representation resulting in the spectrum governed by Gell Mann Okubo mass relations. For octet baryons:
\beq
M_{8}(I, Y) = M_8^0 - bY +c(I(I+1)-Y^2/4)
\eeq
and equal spacing of levels for both $\{10\}$ and $\{\overline{10}\}$ states:
\beqa
M_{10}(Y) &=& M_{10}^0 - aY,  \\
M_{\overline{10}}(Y) &=& M_{\overline{10}}^0 - a'Y.
\eeqa

A second related consequence of symmetry breaking is the mixing of various states
with  the same combination of ($I$, $Y$) states among different $SU(3)$
representations. In particular we expect $N_8$ and $N_{\overline{10}}$ will mix to
  yield $N(939)$ and some $N^*$ state. Similarly $\Sigma_8$ and $\Sigma_{\overline{10}}$ will mix. These mixings will induce a shift in masses from the above octet
  symmetry breaking relations for the mass eigenstates as well as cause 
  suppression or enhancement of the decay amplitudes. 
\vskip0.3cm
\subsection{Mixing angle from the spectrum}

\vskip0.2cm

If we identify $N_8$, $\Lambda (1115)$, $\Sigma_8$, $\Xi (1318)$ as octet
states and $\Theta^+(1540)$, $N_{\overline{10}}$, $\Sigma_{\overline{10}}$ and
$\Xi_{3/2}(1862)$ as antidecouplet states and assign $ N(939)$ and $N(1710)$
as nucleon-like mass eigenstates and similarly $\Sigma (1193)$ and  $\Sigma
(1880)$ as $\Sigma$ -like states, we may find mixing angles $\theta$ as
follows:

\beqa
|N(939)> &=&  cos\  \theta_N |N_8> -  sin\ \theta_N |N_{\overline{10}}>\\
|N(1710)> &=&  sin\ \theta_N |N_8> +  cos\ \theta_N |N_{\overline{10}}>
\eeqa
and 
\beqa
|\Sigma(1193)> &=& cos\ \theta_{\Sigma} |\Sigma_{8}> - sin\  \theta_{\Sigma}
|\Sigma_{\overline{10}}> \\
|\Sigma(1880)> &=& sin\ \theta_{\Sigma} |\Sigma_{8}> + cos\  \theta_{\Sigma}
|\Sigma_{\overline{10}}>.
\eeqa

It is easily obtained that 
\beqa
939 +1710 &=& <N_8|H|N_8> +<N_{\overline{10}}|H|N_{\overline{10}}>\\
&=& M_8^0-b+{c \ov 2} + M_{\overline{10}}^0-a'\\
&=& 1115 +1755 -a' -b +{c \ov 2}.
\eeqa

From the masses of $\Theta^+$ and $\Xi_{3/2},$ we get $a'=107$ MeV, yielding
$b-{c \ov 2} =114 $ MeV. Using the masses of $\Lambda(1115)$ and $\Xi(1318)$ we
find $b+{c \ov 2} = 203$ MeV. We further have 
\beqa
cos\ 2 \theta_N(1710 -939)& = &<N_{\overline{10}}|H|N_{\overline{10}}> - <N_8|H|N_8>\\
 &=& M_{\overline{10}} - M_8 -a' +b - {c \ov 2}\\
&=& 647 \  \mbox{MeV}
\eeqa

This implies that $cos\  2 \theta_N = 0.84$, which translates into
$tan^2\ \theta_N =0.087$. 

If we denote $<N_8|H|N_{\overline{10}}>= <\Sigma_8|H|\Sigma_{\overline{10}}> = \delta$, we
  will have 
$$
2\delta = (M_{10}-M_8-a+(b-{c \ov 2}))\ tan\ \theta_N
$$
yielding a value $\delta  = 220$ MeV as signifying the extent of representation mixing. 

We may carry out a similar analysis in the mass mixing in the $\Sigma$ - sector
to find

\beqa
1880 + 1193 &=& 1755 +1115 + 2c\\
cos\  2\theta_\Sigma\  (1880 - 1193) &=& 1755 - 1115 - 2c. 
\eeqa

We obtain $cos\ 2\theta_\Sigma = 0.63 $ and corresponds to the representation
mixing value of $\delta$ =264 MeV instead.

As a result of the mixing the expectation value of the $N$ and $\Sigma$ in
the octet representation will be $M_{N_{8}} = M_8-b+ {c \ov 2} = 1001$ MeV
and $M_{\Sigma_8} = M_8 + 2c = 1293$ MeV. Gell Mann Okubo relation for the octet
will imply 
$$
2 M_{N_8} - M_{\Sigma_8} = 3 M_{\Lambda} - 2 M_{\Xi} = 709 \ MeV
$$

Before mixing is applied the left hand side of the expression will have $2\times 939
-1193 = 645$ MeV. When mixing is taken into account it is instead $2\times 1001$ -1293
= 709 MeV, much more precise. Indeed the nucleon (and $\Sigma$) needs $\{\overline{10}\}$
admixture to fit GMO formula better!{\footnote {However the Guadagnini hybrid mass relation $8M_{N_8}+3M_{\Sigma_8}=11M_{\Lambda}+8M_{\Sigma^*}-8M_{\Xi^*}$, which is satisfied to an amazing $0.1\%$ before mixing loses a shade when mixing is taken into account to an agreement at about $2 \%$ level.}} 
\vskip0.3cm
\subsection{Mixing angle from the decays}

\vskip0.2cm

The mixing angle can also be deduced by studying the decay width of the
states. The couplings are assumed to be preserving $SU(3)$, but in the
computation of the width, the phase space is computed with the actual [$SU(3)$
  broken] mass values. We have already observed that while for $\{10\}_{{3/2}^
+} \to
\{8\}_{{1/2}^+} + \{8\}_{0^-}$, the rates are proportional to $(G_0+{1 \ov 2}
G_1)^2$, the transition rates for $\{\overline{10}\}_{{1/2}^+} \to
\{8\}_{{1/2}^+} + \{8\}_{0^-}$ are governed by the factor $(G_0 - G_1
-{1 \ov 2} G_2)^2$. The partial decay widths for the antidecouplet baryons are
given by 

\beqa
\Gamma(\Theta^+ \to KN) &=& {3 \ov 2\pi} {(G_0 - G_1
-{1 \ov 2} G_2)^2 \ov (M_N +M_{\Theta})^2} {M_N \ov M_{\Theta}} p^3_{\Theta \to
    KN} \left|\pmatrix{ 8&8&\overline{10}\cr K&N&\Theta}\right|^2\\
  \Gamma(\Xi^{--} \to \pi^-\Xi^-) &=& {3 \ov 2\pi} {(G_0 - G_1
-{1 \ov 2} G_2)^2 \ov (M_{\Xi^{--}} +M_{\Xi^{-}})^2} {M_{\Xi^{-}} \ov M_{\Xi^{--}}} p^3_{\Xi^{--} \to
    \pi^- \Xi^-} \left|\pmatrix{ 8&8&\overline{10}\cr \pi^-&\Xi^-&\Xi^{--}}\right|^2  
\eeqa
with 
$$
p_{B_1 \to B_2 M} = \sqrt{(M_{B_1}^2 - (M_{B_2} + M_M)^2) (M_{B_1}^2 -
  (M_{B_2} - M_M)^2)} /2 M_{B_1}.
$$

Notice that there is no $SU(3)$ symmetric coupling that will permit
$N_{\overline{10}} \to \Delta + \pi $ (because $\{\overline{10}\} \{10\}\{8\}$
coupling does not exist.)
However, since $N(1710) \to \Delta(1232) \pi$ has a partial width of about 5
MeV, this is a clear indication that this state can not be a pure
antidecouplet. It is the octet part of the state that is responsible for the
$\Delta \pi $ decay mode. Comparison of the partial width for $N(1710) \to
\Delta \pi$ with that of $\Delta \to N \pi$ will give us the measure of the
admixture. Since $|N(1710> = cos\ \theta_N |N_{\overline{10}}> + sin\ \theta_N
|N_8>$, we get 
$$
\Gamma(N(1710)\to \Delta \pi) = sin^2\ \theta_N {3 \ov 2\pi} {(G_0+ {1 \ov
    2}G_1)^2 \ov (1710
  +1232)^2} {1710 \ov 1232} p^3_{N^*\to\Delta\pi} {4 \ov 5} \sim 5 MeV. 
$$
Comparing this with   
$$
\Gamma(\Delta(1232) \to N \pi) = cos^2 \ \theta_N {3 \ov 2\pi}
      {(G_0+{1 \ov 2}G_1)^2 \ov (939
  +1232)^2} {1232 \ov 939} p^3_{\Delta \to N\pi} {1 \ov 5} =120 MeV,
 $$
we find that $tan^2\ \theta_N = 0.0035$. Note a very small admixture is all
that is needed to get a 5 MeV partial decay width for $N(1710)\to \Delta \pi$. 

We conclude that mixing angle from both mass spectrum and decay data are reasonably small. The decay amplitudes appear to give a much smaller value of mixing
parameter than the data on the mass spectrum of states. 

It will be useful to contrast these results with other analysis. Pakvasa and Suziki \cite{pakvasa}, who assumed the mixing octet to consist of
N(1440), the old Roper resonance and $\Sigma(1660)$ instead of $N(939)$ and
$\Sigma(1193)$ along with appropriate excited states
for both $\Lambda$ and $\Xi$, find an even larger discrepancy in view of their choice for states. Since they use Roper resonance in
place of the nucleon of the ground state baryon octet,they get  a substantial mixing from the mass spectra
and therefore can not reconcile with a much smaller mixing angle that is
indicated  from the decay
amplitudes. Weigel \cite{weigel}, who has made an extensive study of the
spectrum of Skyrmion states, identifies such an octet state with the radially
excited Skyrmion. It is not clear to us why $N_{\overline{10}}$ would prefer to mix
with the radially excited state over the ground state. Perhaps we need to
compute mixing with all the three levels. However in such an analysis, in the
leading order, we expect the radially excited state to be orthogonal to the
ground state and hence should not be giving any qualitatively different
answers.   

Mixing of states may arise also with higher iso-rotational levels in the
Skyrmion  sector, that we have listed. Analysis have been carried out by
including $\{27\}_{1/2}$ and $\{\overline{35}\}_{1/2}$ (which are expected to be much
      and very much heavier respectively) for $J={1 \ov 2}$ states and $\{27\}_{3/2}$,
     $ \{35\}_{3/2}$ states with  $\{10\}_{3/2}$ baryons for $J={3 \ov 2}$
      levels. 

We don't have much experimental support for higher states of baryons as of
now and so such an analysis is of academic interest only. The basic premise that $\{\overline{10}\}$ states have a small admixture of the
ground state octet baryons appears to be born out by experimental data so far
available. while mixing angles in the most basic interpretation are indeed
small, the decay data appears to suggest much smaller mixing. We will see that
an alternative explanation in terms of Constituent quark model that will keep
track of the number of strange quarks in a state will imply substantial mixing
of representations, when $SU(3)_F$ is broken. 
\vskip0.5cm
\section{Constituent Quark Model}

\vskip0.3cm
In constituent quark model framework, the hadrons are considered built up of dressed
(valence) quarks much the same way nuclei are built using nucleons and model
effective interaction among dressed quarks. In a naive uncorrelated quark
model for pentaquark baryons, we expect the ground state to be a $J^P = 1/2^-$
state with flavour quantum number to be both $\{\overline{10}\}_F$ and
$\{8\}_F$. They expect to be accompanied with states with $J^P=3/2^-$ with
roughly the same difference in mass as between $N$ and $\Delta$. Jaffe and
Wilczek argue that there is a lower energy state with positive parity,
exploiting a strong correlation among quarks that picks diquark pairs that are
antisymmetric in color, spin and flavour. Call these condensates ${\bf Q}$ that are 
$\{\overline{3}\}_C$ and  $\{\overline{3}\}_F$ scalars. Pentaquark baryons are then ${\bf QQ}{\bar q}$ composites. For two such diquarks,
along with an antiquark, to form a color singlet they must be antisymmetric in
color. In order to form a $\{\overline{10}\}_F$, the diquark pair have to be in a
flavour symmetric $\{\overline{6}\}_F$, which in turn implies orbital excitation
and $l=1$. When combined with an antiquark, we have the pentaquark states in
$\{\overline{10}\}_F$ and $\{8\}_F$ , expected to be degenerate, both having
$J^P=1/2^+$ and $3/2^+$, with the spin orbit coupling providing the splitting
as found in the separation between $N$ and $\Delta$. The nucleon like states
and  the $\Sigma$ -like states of the $\{\overline{10}\}$ and $\{8\}$, when
subjected to $SU(3)$ breaking are expected to mix {\it ideally}, so that the
lighter member has no strange quarks and the higher has a $s\overline{s}$ pair for
the nucleon-like member. This is similar to the vector meson nonet with $\omega - \phi$ mixing, where $|\omega> ={1 \ov \sqrt{2}}(|u{\bar u}> -|d{\bar d}>)$ and $|\phi> = |s{\bar s}>$. Ideally mixed nucleon like states then are:
 
\beqa
|N_{ud}> &=& \sqrt{2/3}\ |N_8> + \sqrt{1/3}\ |N_{\overline{10}}>\\
|N_s>    &=& -\sqrt{1/3}\ |N_8> + \sqrt{2/3}\ |N_{\overline{10}}>
\eeqa
with $|p_{ud}> = |udud\overline{d}>$, $ |n_{ud}> = |udud\overline{u}>$, $|p_s>=
|udus\overline{s}>$ and $|n_s> = |udsd\overline{s}>$. For ideal mixing $tan\ \theta =
\sqrt{2}$.  

Similarly the $\Sigma$ -like states are, indeed: 
\beqa
|\Sigma_{ud}> &=& \sqrt{2/3}\ |\Sigma_8> + \sqrt{1/3}\ |\Sigma_{\overline{10}}>\\
|\Sigma_s>    &=& -\sqrt{1/3}\ |\Sigma_8> + \sqrt{2/3}\ |\Sigma_{\overline{10}}>
\eeqa  
 with $|\Sigma^+_{ud}> = |udus\overline{d}>$, $|\Sigma^+_s>= |usus\overline{s}>$;
 $|\Sigma^0_{ud}> = {1 \ov \sqrt{2}}
 |udus\overline{u}> + {1 \ov \sqrt{2}} |udds\overline{d}>$, $|\Sigma^0_s> = |usds\overline{s}>$; and  $|\Sigma^-_{ud}> =
 |udds\overline{u}>$, $|\Sigma^-_s> = |dsds\overline{s}>$.

Jaffe and Wilczek expect the mass spectrum to be governed by the number of $s$
quarks and $\overline{s}$ antiquark the baryon has. If $H=M_0+ (n_s+n_{\overline{s}})\alpha
+n_s \beta$, it will imply ideal mixing of $\{\overline{10}\}$ and $\{8\}$. We will
then have the sequence  of pentaquark baryons with $N^*_{ud}\  < \Theta^+\ 
< \Sigma^*_{ud}, \Lambda^* \ < \ N^*_s \ < \ \Xi^*_{1/2},\ \Xi^*_{3/2} \ < \
\Sigma^*_s$. The states identified by them to fit this sequence, apart from
$\Theta^+(1540)$, are $N^*_{ud}(1440)$, $N^*_s(1710)$, $\Lambda^*(1600)$,
$\Sigma^*_s(1880)$. They agree with the above sequence if we accept their
analysis of the systematics for exotic $\Xi$ decays, where they argue that
there are nearly degenerate $\Xi_{1/2}$ and $\Xi_{3/2}$ in 1855 - 1860 mass
region\cite{jw2}.  

This large mixing angle is incompatible with the information from decay data.
In particular, $\Gamma(N(1710) \to \Delta \pi)$ will turn out to be absurdly
large, if $tan\ \theta = \sqrt{2}$. Further the narrow width for $\Theta^+$
and $\Xi^{--}_{3/2}$ will be in sharp conflict with the expected large width for
$N_{ud}$ and $N_s$. While the broad Roper resonance fits the bill for
$N_{ud}$, it is not clear whether there is another broad nucleon-like state in
that region instead, if $N(1710)$ is to be discounted as the candidate for
$N_S^*$. 

Equally dramatic is the prediction that both $\{8\}_F$ and $\{\overline{10}\}_F$
will be nearly degenerate before $SU(3)_F$ is broken. This is indeed the
argument for interpreting that the $\Xi$ resonances cover both $I = 1/2$ and
$I = 3/2$ components in the same region.

In contrast against the Chiral Soliton Model, the Pentaquark model predicts in
addition to $J^P = 1/2^+$ state, $J^P = 3/2^+$ states as well with an expected
mass difference, which should be of the same order as $(M_{\Delta} - M_N)$. We
expect many more baryonic states than what has been reported. We observe that
there is no interlocking of spin and isospin that was characteristic in the
soliton sector.

We need more definitive experimental evidence before we are able to rule in
favour of 
either Chiral Soliton Model or any of the specific constrained Constituent
Quark Models. There are other variants of Jaffe-Wilczek diquark correlations;
for example the one by Lipkin and Karliner \cite{lipkin} postulates two
separate clusters one made up of the same diquark and another ($qq\overline{q}$)
cluster in which the two quarks are in color symmetric $\{6\}_C$ state.The observed narrow width for $\Theta^+$ is attributed to the fact that the
clusters are kept apart due to angular momentum barrier.

\section{Comparisons and conclusion}

Both in chiral model as well as in constituent pentaquark models with diquark
correlations folded in, the spin parity assignment favoured is $J^P ={1 \ov
  2}^+$. However, while all rotational excitations in chiral soliton sector
are necessarily  of  positive parity, there is no reason to exclude negative
parity baryons in  the pentaquark picture. The additional states in the CSM
are are attributed to radial excitations, all of which will have positive
parity, and same $SU(3)_F$ quantum numbers. In contrast, in CPQM we expect a
more or less degenerate octet of states in addition to $\{{\bar 10}\}$
baryons, both of spin 1/2 and 3/2 variety  and further (perhaps a bit more
massive ) negative parity states. More detailed spectroscopy in this mass
region can clarify whether such additional states are present.      

In the CSM, Nucleon like states in the ground state octet and the exotic
antidecouplet will indeed mix. These mixing angles remain small and generally
found to be so with our assignment, which parallels Diakanov et al's
choice. Even so, the mixing angle as obtained from the decay widths appear
further smaller than that obtained from mass spectrum. In contrast,
when $SU(3)_F$ is broken in the CPQM we expect a large mixing of the nearly
degenerate $\{8\}$ and$ \{\bar {10}\}$ multiplets to give it the nature of
ideal mixing.  This will lead to a large strangeness content of one of the
nucleon like states, say $N(1710)$  and so it must have a significant
branching ratio into $\Lambda K$ and $\Sigma K$ channels in order that Zweig
rule is obeyed. Further since  it is made up of a substantial component of
octet, the coupling to baryons and mesons such as $\Delta \pi$, $\Lambda K$
etc. will be comparable to the strength of $\Delta NK$ coupling. This is in
conflict with the observed branching ratio and small widths of this
state. Admittedly more detailed analysis is called for before we can confirm
$N(1710)$ as the candidate $uuds{\overline s}$ state. 

Is there a deeper reason for the dramatic narrow width? We note that the
degenerate octet and decouplet states arise from $(\bar 3,\bar 6) \oplus (\bar
6,\bar 3)$ of the underlying $SU(3)_L \times SU(3)_R$ symmetry. In the same
scheme baryon octet belongs to $(1,8) \oplus (8,1)$. In the limit of exact
chiral symmetry (left handed currents transforming like $(8,1)$ and right
handed currents as $(1,8)$ ) there will be no coupling between pentaquark
baryons and the usual triquark octet. Thus $\Theta^+ \to NK$ decays are
inhibited on account of the underlying chiral symmetry\cite{ioffe}. In
exact chiral symmetry, (limit in which pion masses vanish) we expect stable
$\{\bar {10}\}$ baryon states\cite{beane}. 

Both view points have room for many additional states; in CSM radial
excitations will give further states with same flavour quantum numbers and in
CPQM many other permutations of correlated clusters are possible as well. Of
course several of these features could be consequences of higher (${1/N_c}
$) order and hence may not be very  reliable predictions of CSM. Future experiments\cite{futurexp} should nail many of the predictions of both pictures.  
           
  It is satisfying to note, that the chiral soliton model, that starts from an
  underlying chiral symmetry has dynamical ingredients to account for its
  narrow width. In the CPQM, in contrast, small width is due to
  non-overlapping clusters on account of centrifugal barrier. Perhaps we need
  some ingredient that signals approximate chiral symmetry in the CPQM to
  reflect more similarity with the soliton picture. This calls for the
  possibility that we may be able to describe hybrid models that have features
  of both Chiral Soliton picture on the one hand, while being legitimately
  pentaquark constituent structures in terms of relevant variables. Main
  reason for such a prospect has to do with the feature that the narrow width
  is very likely a consequence of underlying approximate chiral symmetry.

\end{document}